# Simulation of Ge on Si Photodiode with photon-trapping micro-nano holes with -3dB bandwidth of >60 GHz at NIR wavelength

**Ekaterina Ponizovskaya Devine[a,b*] , Toshishige Yamada [b,c], Shih-Yuan Wang[b], and M Saif Islam,[a]**

[a]University of California, Davis, CA, 95616, USA
[b]W&WSence Devices Inc*, 4546 El Camino, Los Altos, CA 94022 USA*
[c]University of California, Santa Cruz, Santa Cruz, CA 95064 USA

**Abstract**. The study proposes an ultra-thin back side illuminated (BSI) and top-illuminated, Ge on Si photodetector (PD), for 1 to 1.4 microns wavelength range. The Ge thickness of 350 nm allows us to achieve high-speed performance at >60 GHz, while the nanostructure at the bottom of the Ge layer helps to increase the optical absorption efficiency to above 80%. The BSI PD allows the PD or PD array wafer to be stacked with an electronic wafer for signal processing and transmission for optical interconnect applications such as short-reach links in data centers. Nano-microhole parameters in randomized composite formation on the bottom layer are optimized with Monte-Carlo molecular dynamics simulations incorporating charge transport to enable wide-spectral, highly efficient, and ultra-fast PDs.

*Ekaterina Ponizovskaya-Devine**, E-mail: eponizovskayadevine@ucdavis.edu

## 1 Introduction

The demand for efficient and high-speed vertically illuminated photodetectors (PD) for data centers, lidar applications, and imaging [1,2] stimulates the research that provides new approaches such as photon-trapping for photodetectors [3] and for image sensors [4]. While Silicon photonics produces compact, multifunctional, and low-power photonic circuits built on the same chip, many applications need near-infrared wavelengths that can be achieved with group IV elements. Silicon-germanium (Si–Ge) photodetectors [5–8] are filling these needs in optical communications. Recently, progress in more efficient and fast photodetectors was achieved using the micro-nano-structures in Si and Ge, as well as in III-V semiconductor materials [9] for short [10] and mid-infrared [11] wavelengths. The interest in the Ge PDs is still high as they are approachable solutions for near-infrared and applications that need eye safety. The ultra-thin active layer in PD drastically increases its speed for the vertically illuminated device, as the electrons and holes have a much shorter path.

It was shown in [3] that the micro-holes on the top of the device reduce the reflection and produce the modes that stay in the active layer longer, increasing the photon absorption many times. We propose the Ge on Si structure with the holes on the bottom of the Ge active layer. The fabrication of the structure is possible using the stack CMOS technology [12]. The design is beneficial for the electronics stackable on the same wafer. The upper layer is Si with an antireflection structure.

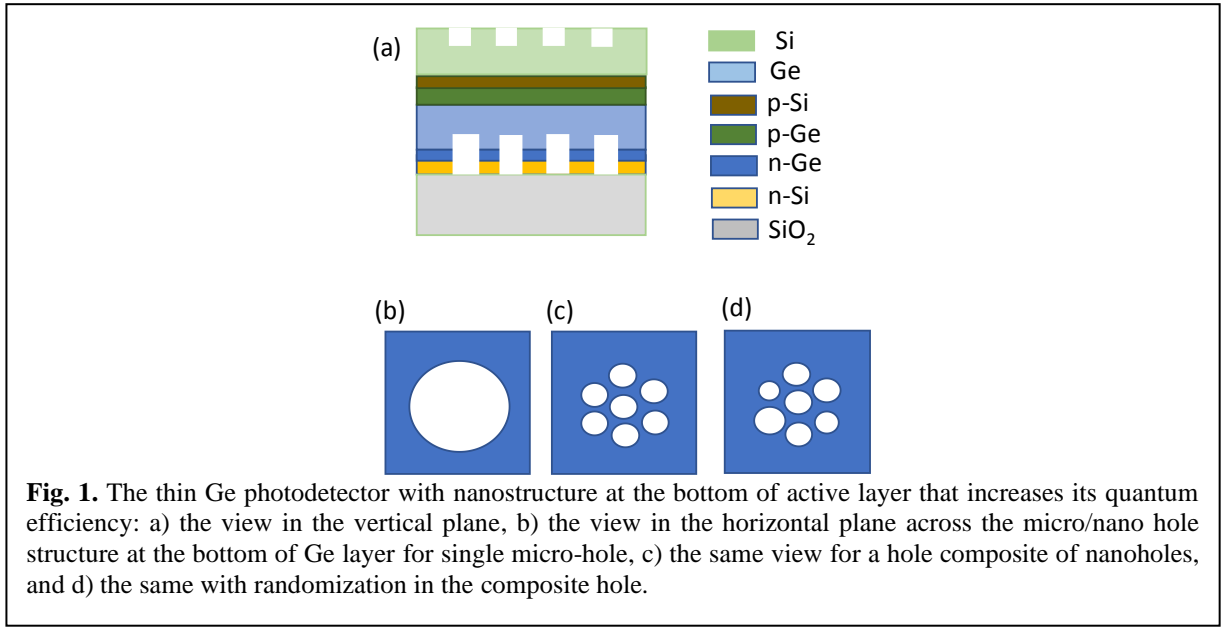

**Fig. 1.** The thin Ge photodetector with nanostructure at the bottom of active layer that increases its quantum efficiency: a) the view in the vertical plane, b) the view in the horizontal plane across the micro/nano hole structure at the bottom of Ge layer for single micro-hole, c) the same view for a hole composite of nanoholes, and d) the same with randomization in the composite hole.

Fig. 1 shows the schematics of the PD. The upper Si layer is 100nm, 100nm p-doped Ge, active Ge layer is only 350nm, next is 100 nm n-doped Ge and 25nm n-doped Si, with the holes through the Si and Ge layer. All the PD is set on the layer of SiO2, see Fig 1a.

We studied a variety of micro-hole configurations (Fig.1b): the cylindrical holes of different sizes from 200nm to 1000nm, period from 300-1200nm and depth from 50-200nm, and the composite holes that form a cluster of 7 holes that forms a larger hole. We also simulated a randomized hole structure that shows a smoother spectrum in a wider range. We used Finite-Difference Time-Domain (FDTD) simulation for absorption and Monte-Carlo simulations for the 3-D electron-hole transport with micro-holes, to reconstruct a pulse and estimate a bandwidth. The 3D electro-hole transport simulations with finite difference algorithms are considerably time-consuming [13]. Our Monte-Carlo simulations were validated on the experimental results from [5].

## 2 Simulations Methodology

Two important factors in the PD design are the optical absorption enhanced by the nanostructure and the electric pulse response defined by the electron-hole transport. We modeled both effects. The absorption was modeled by the FDTD method [14]. The method solves Maxwell's equations for the electric (E) and magnetic (H) components of light, which represent Ampere's and Faraday's laws.

$$\nabla \times H = \epsilon(r)\epsilon_0 \frac{dE}{dt}$$

$$\nabla \times E = -\mu\mu_0 \frac{dH}{dt}$$

$\epsilon$ is the dielectric permittivity that depends on coordinate r and consists of real and imaginary parts that represent material conductivity and are responsible for the light absorption, μ is magnetic permeability that is constant in our case, *t*- time. The dielectric permittivity dependence on the wavelength of the photons is modeled by a set of Lorentz oscillators [14] to match the Ge characteristics.

By calculating the integral Pointing vector that enters the volume and leaves the volume, the absorbed light is calculated [4]. The incident light was perpendicular to the surface. The Perfect Matching Layers (PML) were used at the boundaries perpendicular to the light incident direction and periodical boundaries were applied to the other's boundaries. For randomized micro-holes, a large area of 5x5 unit cells of randomized holes was chosen. FDTD simulation also gives light intensity distribution in the area.

We used the Monte Carlo molecular dynamics method to calculate the pulse response. There are recently several technics that were used for simulations. Device simulation technology for computer-aided design (TCAD) provides a basis for device modeling by using compact behavioral models and sub-circuits relevant to circuit simulation realized in commercial packages and drift-diffusion models [15]. As far as we know, for the circuit characteristics in the presence of micro-holes, we need a method that will be based on the physics of the transport that can be described by the Boltzmann transport equation [16] and the statistical distribution of the electrons and holes.

The commercial software Lumerical Charge [17] offers finite element methods to solve the partial differential equations for electron-hole concentration. A recent study [13] showed that electron-hole transport in 3D simulation is a difficult problem and the finite-element methods could be time-consuming. Instead, the Monte Carlo method offers the microscopic simulation of the motion of individual particles in the presence of the external fields and the internal fields related to the crystal lattice and other charges. The method was widely used for molecular dynamics and the thermodynamic properties of liquids and gases. In semiconductors, transport is dominated by random scattering events due to impurities and lattice vibrations, that randomize the momentum and energy of charged particles in time, so the stochastic techniques to model random scattering events are applicable.

The Monte Carlo technique considers random stochastic events in periodic boundary conditions. This is a challenge in programming, and Yamada and Ferry developed a method based on the Ewalt sum, where the Monte-Carlo code was validated against experimental data from the studies [3,5]. In the Monte-Carlo molecular dynamics method, we generate a random electron-hole pares with probability proportional to the light intensity that was simulated by FDTD and depends on coordinates. Each hole moves to distance *r=r₀ ln(rand)* where *r₀* is the average distance between scattering events that is the product of thermal velocity to the average scattering time and *rand* is a random number in the interval [0,1]. The carrier current density for every coordinate *r* the sum of diffusion, drive, and displacement current

$$j = j_{diff} + j_{dr} + j_{disp} = qD_e\nabla n + qD_p\nabla p + q\text{n}\mu_e E_s + q\text{p}\mu_p E_s + \epsilon_0\epsilon dE_s/dt$$

Thus, the velocity of the electrons and the holes have three components – one is the random scattering that represents diffusion current $j_{diff}$, another is determined by the electric field applied to the PD that represents drift current ($j_{dr}$) and the third is the velocity in the potential charges created by the electron-hole distribution, which represents the displacement current $j_{disp}$. The velocity overall is limited by the thermal velocity. Here *q* is the electron charge, n and p are the electrons and holes concentration, *E_s* is the electrostatic field produced by the voltage, *D_e* and *D_p* are the electrons and holes diffusion coefficients, $\mu_e$ and $\mu_p$ are the electrons and holes mobilities. Instead of solving the finite-element equation for the concentration of electrons and holes in time that has instabilities and is time-consuming, we update the concentration of the carriers tracking the Monte-Carlo motion of super-particles that represents a number of individual carriers.

The idea is to use large units of charges, superparticles, and update the superparticle locations after successive evaluation of Coulomb forces [18]. This is the spirit of Monte-Carlo molecular dynamics giving a great numerical stability [18, 19, 20]. The Monte-Carlo molecular dynamics is discussed in detail from the view point of thermodynamics, and it would be a great interest for materials scientists and chemical engineers [21].

The potential (ϕ) distribution of charges and electrical field with the presence of the micro holes is simulated by solving the Poisson equation.

$$\Delta \varphi = \epsilon \epsilon_0 q (n + N - p - P)$$

$$\nabla \varphi = E_s$$

The boundary conditions are that the potential is equal to voltage *V* at the electrodes and $\nabla \varphi = 0$ at the other boundaries, *n* and *p* are the concentration and *N* and *P* are the doping concentration and is the electrostatic field created by the voltage *V*.

Besides the scattering event, the particle can experience a recombination event with probability proportional to Shockley-Read-Hall recombination [13, 22, 23, 24, 25] and Auger recombination [13, 22, 23, 24, 26], as well as the surface recombination [27] if the particle reached the surface without reaching the electrodes. In this case, the particle will disappear with the probability proportional to the recombination rate.

When the electron or hole reaches the electrode it is calculated as current at the electrode. Initial light could be modeled as a pulse and in this case, we can register a current that corresponds to the pulse. The speed of the response can be characterized by the width of the response pulse or by the width at -3 dB of the Fourier transform of the pulse that will represent the bandwidth. Bandwidth also takes into account the response tail, and, thus, is a better parameter.

## 3   Results and Discussion

We optimized the photodetector for the wide wavelength range between 1 and 1.4 microns. It was observed in [6] that the optimal micro-holes for trapping photons in the Ge layer were 1000-1200nm and the period 1200-1500nm, while smaller holes produced sharp guided resonance. However, the combination of several resonances that are close to each other is possible and can produce a strong absorption in a wide range of wavelengths. We also studied the option to use composite holes that combine the smaller holes into a structure with a diameter of 1000-1200nm. The Monte-Carlo simulations showed that the micro-hole shape can influence the electron-hole transport and, thus, the speed of the device. One of the problems is surface recombination but it can be mitigated with the proper passivation. We position our holes in the bottom filled with silicon oxide that provides good passivation. That allows us to assume that the surface recombination is small. The second problem is that the holes represent an obstacle between the electrodes and the carriers can be stuck in the areas near the bottom of the micro-holes creating charges that slow down the current and result in the tail. Eventually, the carriers will either be recombined or will move out of the area due to diffusion. Reducing the micro-hole size reduces the area of slow down for the carriers. The third issue is the doping and width of the n and p areas. We have seen that relatively small and high-doped n and p areas provide a stronger field and decrease the time of the response. The areas are thin with doping concentration $10^{19} cm^{-3}$.

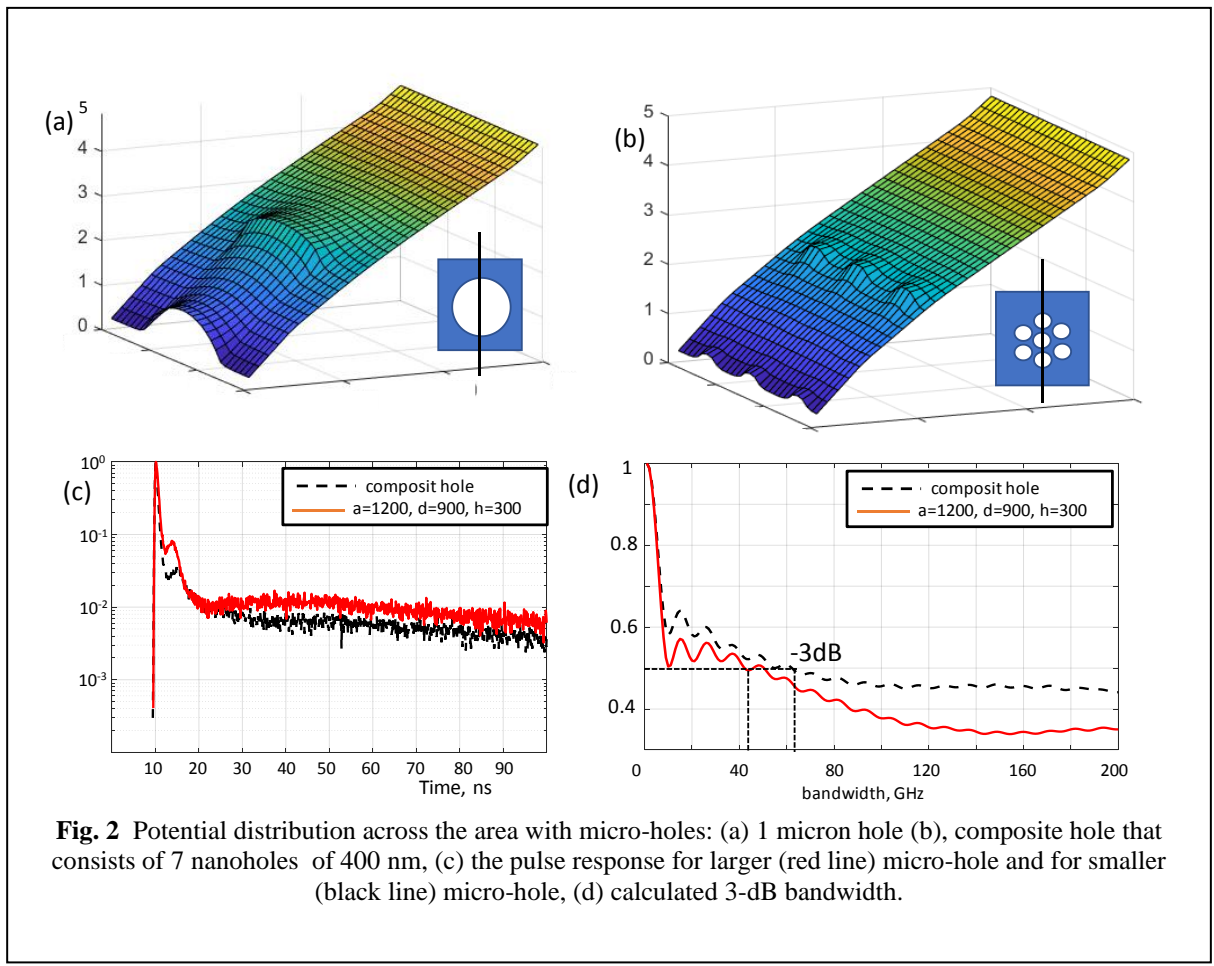

**Fig. 2** Potential distribution across the area with micro-holes: (a) 1 micron hole (b), composite hole that consists of 7 nanoholes of 400 nm, (c) the pulse response for larger (red line) micro-hole and for smaller (black line) micro-hole, (d) calculated 3-dB bandwidth.

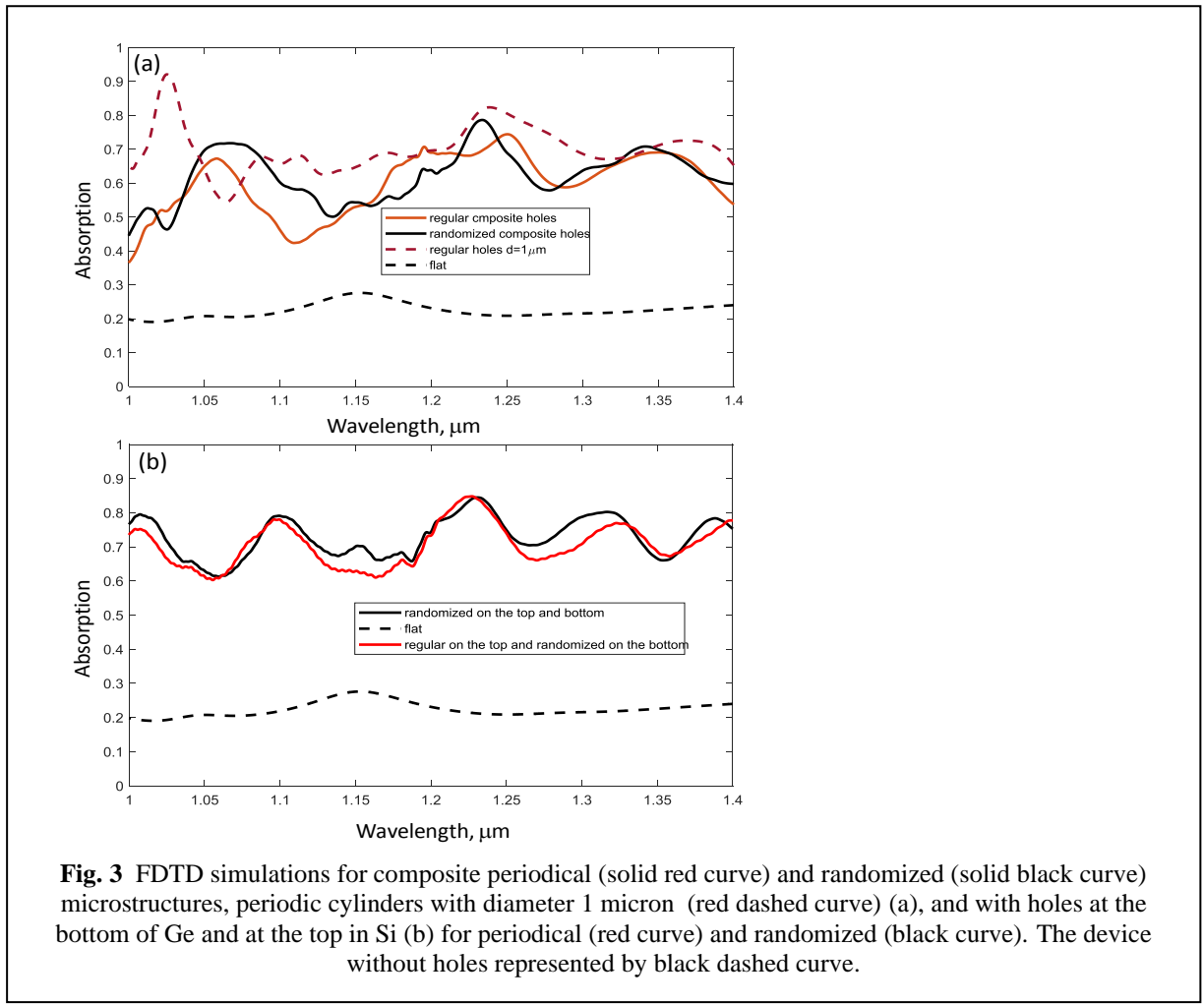

**Fig. 3** FDTD simulations for composite periodical (solid red curve) and randomized (solid black curve) microstructures, periodic cylinders with diameter 1 micron (red dashed curve) (a), and with holes at the bottom of Ge and at the top in Si (b) for periodical (red curve) and randomized (black curve). The device without holes represented by black dashed curve.

Fig. 2 shows the potential $\varphi$ around larger (a) and smaller (b) microholes that are formed by the 5 V voltage. We can see that under the bottom of the holes, there is an area with a lower electric field. As a result, we can see the tail in pulse response, which is shown for the larger and smaller micro-holes in Fig. 2c in the logarithmic scale. The tail is higher for the larger hole. Even though the width of the pulse is short due to the active layer being thin, the tail influences the bandwidth, as shown in Fig. 2d. The bandwidth at -3dB for the composite micro-holes is at 60 GHz, while for the big hole is only about 40 GHz.

The earlier simulations for Ge [6] showed that the optical absorption can be strongly increased by micro-holes with a diameter of 1000-1200nn. However, to reduce the tail we would like to use smaller holes that compose a structure of a diameter of about 1200 nm (Fig.1c). We used the 300 nm diameter nanoholes with a distance of 400nm between the centers.

The results are represented in Fig. 3a for the regular holes on the bottom of Ge (red line) and the structures composed from randomized holes (Fig. 1d) with a diameter of 300 +/- 100 nm and a distance of 400+/-50 nm. The composite nano-holes–produce the absorption comparable with optimal cylindrical micromoles on the bottom. The results were compared with a flat device (black dashed line). The improvement in all cases is significant however the randomization helped to smooth out the resonances. Even better results are achieved by using microstructure in the Si layer on the top to minimize the reflection. The randomization of the composite micro-holes at the top layer did not produce a large improvement, while both cases produced absorption between 60-80% that is better than 1-micron periodic micromoles for most of the wavelengths between 1 and 1.4 microns.

## 4    Conclusions

We have shown that the thin Ge photodetector can use holes in the bottom of the active layer to improve the quantum efficiency and bandwidth for a wide range of wavelengths. The simulations show that the bandwidth can be improved up to 60 GHz for the vertically illuminated photodetector. The quantum efficiency could reach up to 80 % for the thin Ge PD with a Ge thickness of only 350 nm. Thin doping areas with high doping concentrations also contribute to a high bandwidth. We have shown that micro-hole randomization can provide a wide-spectral high quantum efficiency. In comparison with other approaches that use sharp-guided modes or a few sharp-guided modes close to each other [5], the randomized composite holes provide good quantum absorption in a wider wavelength range.

*Acknowledgments*

This work was supported in part by the S. P. Wang and S. Y. Wang Partnership, Los Altos, CA.

**Caption List**

**Fig. 1** The thin Ge photodetector with nanostructure at the bottom of active layer that increases its quantum efficiency: a) the view in the vertical plane, b) the view in the horizontal plane across the micro/nano hole structure at the bottom of Ge layer for single micro-hole, c) the same view for a hole composite of nanoholes, and d) the same with randomization in the composite hole.

**Fig. 2** Potential distribution across the area with micro-holes: 1 micron hole (a), composite hole that consists of 7 nanoholes of 400nm (b), the pulse response for larger (red line) micro-hole and for smaller (black line) micro-hole (c) the bandwidth (d)

**Fig. 3** FDTD simulations for composite periodical (solid red curve) and randomized (solid black curve) microstructures, periodic cylinders with diameter 1 micron (red dashed curve) (a), and with holes at the bottom of Ge and at the top in Si (b) for periodical (red curve) and randomized (black curve). The device without holes represented by black dashed curve.